# Design of Experiments with Sequential Randomizations on Multiple Timescales: The Hybrid Experimental Design

Inbal Nahum-Shani
Institute for Social Research
University of Michigan
Contact Information: inbal@umich.edu

John J. Dziak
Institute for Health Research and Policy
University of Illinois Chicago

Hanna Venera
School of Public Health and Institute for Social Research
University of Michigan

Angela F. Pfammatter
College of Education, Health, and Human Sciences
The University of Tennessee Knoxville

Bonnie Spring
Feinberg School of Medicine
Northwestern University

Walter Dempsey
School of Public Health and Institute for Social Research
University of Michigan

This work was funded by National Institutes of Health, Grants U01 CA229437, P50 DA054039, R01 DA039901, and R01 DK108678

Correspondence should be sent to: Inbal Nahum-Shani, Institute for Social Research, University of Michigan, inbal@umich.edu

**Abstract**

Psychological interventions, especially those leveraging mobile and wireless technologies, often include multiple components that are delivered and adapted on multiple timescales (e.g., coaching sessions adapted monthly based on clinical progress, combined with motivational messages from a mobile device adapted daily based on the person's daily emotional state). The hybrid experimental design (HED) is a new experimental approach that enables researchers to answer scientific questions about the construction of psychological interventions in which components are delivered and adapted on different timescales. These designs involve sequential randomizations of study participants to intervention components, each at an appropriate timescale (e.g., monthly randomization to different intensities of coaching sessions and daily randomization to different forms of motivational messages). The goal of the current manuscript is twofold. The first is to highlight the flexibility of the HED by conceptualizing this experimental approach as a special form of a factorial design in which different factors are introduced at multiple timescales. We also discuss how the structure of the HED can vary depending on the scientific question(s) motivating the study. The second goal is to explain how data from various types of HEDs can be analyzed to answer a variety of scientific questions about the development of multi-component psychological interventions. For illustration we use a completed HED to inform the development of a technology-based weight loss intervention that integrates components that are delivered and adapted on multiple timescales.

## Introduction

Advances in mobile and wireless technologies offer tremendous opportunities for building psychological interventions that include multiple components, where a component is defined as any aspect of the intervention that can be separated out for investigation (Collins, 2018). In practice, decisions about the delivery of these components are often made on multiple timescales, as illustrated in the following three examples. First, in a technology-based weight loss intervention (Pfammatter et al., 2019; Spring et al., 2020), a decision may be made every several weeks about whether to enhance the intensity of coaching sessions based on information about progress in weight loss. Additionally, every day a decision may be made about whether to deliver a motivational message via a mobile device to encourage self-monitoring of dietary intake, based on information about prior self-monitoring. Second, in an intervention designed to increase engagement in evidence-based tobacco cessation treatments (Fernandez et al., 2020), a decision may be made weekly about whether or not to deliver a text message to encourage engagement in treatment, based on information about prior engagement; and after several months a decision may be made about whether or not to augment the text messages with brief counseling calls, based on information about treatment enrollment by that time. Third, in an intervention for young adults with type 1 diabetes (Stanger et al., 2021), decisions may be made at the beginning of the program about whether or not to deliver two components–incentives to promote consistent daily adherence to goals, and online health coaching to teach effective problem solving—based on baseline diabetes indicators such as duration of diabetes and device use. In addition, decisions may be made daily about what type of feedback to deliver via a mobile device, based on prior adherence to self-monitoring blood glucose.

Investigators often have scientific questions about how to best guide decisions made on various timescales about intervention components. For example, is it better to start with low (vs. moderate) intensity coaching sessions, when motivational text messages are also being delivered daily? Is a daily text message more effective in promoting daily self-monitoring, when more intense coaching is being delivered? What combination of digital intervention components (e.g., mobile app, text messages, non-monetary digital incentives) should be offered initially, if four weeks later coaching sessions will be added for participants not showing sufficient weight loss?

Existing experimental designs can be used to empirically inform decisions about multiple intervention components that are introduced (a) at a single point in time (e.g., standard factorial designs; Collins, 2018); (b) sequentially over time, at relatively slow timescales, such as over several weeks or months (e.g., the Sequential Multiple Assignment Randomized Trial [SMART]; Lavori & Dawson, 2000; Murphy, 2005); or (c) sequentially over time, at relatively fast timescales, such as every day, several hours or minutes (e.g., the Micro-Randomized Trial [MRT]; Qian et al., 2022). However, although each of these experimental approaches is suitable for empirically informing intervention decisions on a particular timescale, they cannot be used to answer scientific questions about multiple intervention decisions that are made on different timescales.

The Hybrid Experimental Design (HED) is a new experimental approach developed explicitly to help investigators answer scientific questions about the selection and integration of intervention components that may be delivered on multiple timescales (Nahum-Shani et al., 2022). The HED involves sequential randomizations on multiple timescales corresponding to the time period at which each component should be considered. The current manuscript conceptualizes the HED as a special form of a factorial design, wherein different factors are introduced on different timescales. We discuss how the flexibility of this experimental approach offers tremendous opportunities to address a variety of scientific questions about which components to deliver and under what conditions, where each component can be delivered on a different suitable timescale. While the HED is highly flexible and can take on many forms depending on the scientific questions motivating the study, we focus here on three types of HEDs: (1) those that integrate a standard factorial design and a SMART design (i.e., hybrid factorial-SMART); (2) those that integrate a standard factorial design and an MRT (i.e., hybrid factorial-MRT); and (3) those that integrate a SMART and an MRT (i.e., hybrid SMART-MRT). We discuss how data from these three types of HEDs can be analyzed to answer different scientific questions about the development of multi-component psychological interventions. Throughout, we use examples based on a completed study to inform the development of a weight loss intervention that integrates components that are delivered and adapted on multiple timescales (Pfammatter et al., 2019; Spring et al., 2020). The details of this study are modified here for illustrative purposes to demonstrate how different scientific questions require a different form of HED.

## Motivation for HEDs

The HED is an experimental approach to help investigators develop psychological interventions with multiple components that are sequenced and adapted on multiple timescales (Nahum-Shani et al., 2022). The term adaptation here refers to the use of dynamic (ongoing) information about the individual to decide which component to deliver and when, with the goal of addressing an individual's changing needs in the course of the intervention. For example, consider a weight loss intervention that begins with a mobile app alone, and then at week 2, if the individual does not lose a sufficient amount of weight the initial intervention is augmented with coaching; otherwise the individual continues with the initial intervention (Pfammatter et al., 2019; Spring et al., 2020). This intervention decision making process involves adaptation because dynamic information about weight loss is used to decide whether or not to augment the mobile app with coaching. This information is often referred to as a "tailoring variable," which is a term used to describe the information used in making intervention decisions (Nahum-Shani et al., 2012a; 2012b). Note that the adaptation in this example occurs after 2 weeks because the goal is to address conditions (insufficient weight loss) that unfold over 2 weeks. The underlying assumption is that individuals who do not lose a sufficient amount of weight by week 2 are unlikely to achieve the long-term goal of the weight loss program (i.e., 5% weight loss by month 6) and hence need more support in the form of coaching in order to succeed (Pfammatter et al., 2019; Spring et al., 2020). Those who lose sufficient weight by week 2 are likely to achieve the long-term goal of the program and hence do not need additional support in order to succeed. For these individuals, providing coaching at week 2 would be unnecessarily costly and burdensome. Hence, the goal of the adaptation is to increase the number of individuals who achieve long-term success (i.e., improving overall effectiveness of the intervention) while minimizing cost and burden (i.e., improving resource-efficiency of the intervention).

Advances in digital technologies offer tremendous opportunities to deliver and adapt interventions in real-time, real-world settings (Koch et al., 2021; Nahum-Shani et al., 2015; Webb & Cohen, 2021). For example, the mobile app used in a weight loss program can leverage information collected via the mobile device to decide whether or not to send supportive prompts (e.g., via a push notification). Suppose that every day around mid-day, if information from the mobile device indicates that the individual did not self-monitor their dietary intake since waking up, a message is sent via the mobile device encouraging the individual to use the app to self-monitor their dietary intake; otherwise no message is sent. In this case, the adaptation occurs on a much faster timescale than the previous example because the goal is to address conditions (insufficient self-monitoring of dietary intake) that change relatively fast (on a daily basis). The underlying assumption is that insufficient self-monitoring on a given day would undermine the formation of a habit (defined as "a motor or cognitive routine that, once it is triggered, completes itself without conscious supervision"; Bernacer & Murillo, 2014) to self-monitor dietary intake and, in turn, undermine the likelihood of achieving the long-term goal of the weight loss program.

Although technology can be used to deliver support in a way that is accessible and relatively inexpensive, insufficient engagement remains a critical barrier to the effectiveness of digital interventions (Nahum-Shani et al., 2022). Human delivery of interventions, such as coaching delivered by clinical staff, can be more engaging (Ritterband et al., 2009; Schueller et al., 2017) but often more expensive and burdensome. Hence, integrating digital and human-delivered intervention components requires balancing effectiveness against scalability and sustainability (Mohr et al., 2011; Schueller et al., 2017; Wentzel et al., 2016). Since digital and human-delivered components are typically delivered and adapted on different timescales—fast (e.g., every day, hour, minute) and slow (e.g., every few weeks or months), respectively—there is increasing need to develop multi-modality adaptive interventions (MADIs)—interventions in which digital and human-delivered intervention components are integrated and adapted on different timescales appropriate for each (Nahum-Shani et al., 2022). The development of MADIs requires answering scientific questions about how to best integrate intervention components that are delivered and adapted on multiple timescales. In the next sections we provide examples of scientific questions about constructing effective MADIs and discuss how HEDs can be used to answer these questions.

## Key Features of HEDs

The HED was designed to help investigators answer scientific questions about the construction of MADIs. The HED can be viewed as a flexible factorial design. A factorial design is a randomized trial involving two or more factors (i.e., independent variables that are manipulated systematically); possible levels of each factor are crossed with levels of the other factors to form a design with multiple experimental conditions to which study participants are randomized (Collins, 2018). The resulting data can be used to estimate the main effect of each factor, as well as interactions between factors. The main effect is defined as the difference between the mean outcome at one level of the factor and the mean outcome at another level, averaging over the levels of all remaining factors (Collins, 2018). Interactions capture the extent that the effect of one factor varies by the levels of other factors. For example, a two-way interaction can be defined as half the difference between two effects: (1) the effect of a particular factor (i.e., the difference in the mean outcome between the two levels of the factor), at one level of the other factor, averaging over the levels of all remaining factors; and (2) the effect of that particular factor, at the second level of the other factor, averaging over the levels of all remaining factors (for details, see Collins, 2018).

The extant literature demonstrates the efficiency of various types of factorial designs in answering scientific questions about the selection of intervention components (Collins, 2018; Nahum-Shani & Dziak, 2018; Nahum-Shani et al., 2018). Specifically, factorial designs enable investigators to combine multiple experimental cells in different ways, such that outcome data from each study participant can be used to test more than one main effect, thereby answering multiple scientific questions about the selection of intervention components. Collins and colleagues (2009) described this as the "recycling" of study participants and discussed the efficiency of this approach for estimating both main effects and interactions.

Building on this literature, the HED can be conceptualized as a flexible factorial design wherein subjects are sequentially randomized to factors on multiple timescales. Each factor corresponds to an intervention component; participants are randomized to each factor at decision points relevant to that component, namely at time points at which decisions should be made in practice about whether and how to deliver this component. Further, each factor is matched with a primary outcome that measures changes at a suitable timescale. The structure of the HED should match the scientific questions of interest and the relevant timescale for each component under investigation.

## HED Examples

In this section, we discuss three types of HEDs that can be used to address different sets of scientific questions. For each HED, we discuss how the data can be analyzed to address the pre-specified scientific questions.

### Hybrid Factorial-SMART

Suppose investigators are motivated to answer the following questions about the development of a weight loss intervention that integrates human-delivered and digitally delivered components. These questions concern two decision points: the beginning of the program and two weeks later. Specifically, at the beginning of the program investigators would like to know if it is beneficial (in terms of weight loss at month 6), (Q1) to offer a mobile app that includes on-demand weight loss strategies and resources (App); and/or (Q2) to offer weekly coaching (Coaching). Further, investigators would like to know (Q3) whether individuals who show early signs of non-response at week 2 would benefit from adding meal replacement (Meal).

To answer these questions, the investigators may consider an experiment with three factors, one factor for each component. Using italicized abbreviations to represent experimental factors,

*App* refers to the factor corresponding to the mobile app, *Coaching* refers to the factor corresponding to weekly coaching sessions, and *Meal* refers to the factor corresponding to meal replacement. Each factor will have two levels: On and Off. Note that in practice, decisions regarding the components App and Coaching should be made at the beginning of the program. Hence, study participants should be randomized to the corresponding factors *App* and *Coaching* before beginning the program. The decision regarding Meal should be made at week 2. Hence, study participants should be randomized to the corresponding factor *Meal* at week 2. Also notice that the question regarding the Meal component only concerns non-responders at week 2. Hence the randomization to the factor *Meal* should be restricted to non-responders; that is, only non-responders at week 2 should be randomized to this factor.

The experimental design in Figure 1a can provide data for addressing questions Q1-Q3. At the beginning of the study, participants are randomized to one of four experimental options resulting from crossing the levels of the two factors *App* and *Coaching*. This is similar to a standard 2×2 factorial experiment. Then, at week 2, non-responders are randomized again to the two levels of *Meal*, whereas responders continue with their assigned initial intervention. Thus, there are three possible experimental options at week 2, two for non-responders and one for responders. Overall this design integrates a standard 2×2 factorial experiment with a Sequential, Multiple Assignment, Randomized trial (SMART; Lavori & Dawson, 2000; Murphy, 2005).

A SMART is itself already a form of factorial design (Nahum-Shani et al., 2012a) involving sequential randomizations to experimental factors. In a prototypical SMART, there are two decision points at which individuals are randomly assigned to factors, but random assignment at the second decision point is restricted to non-responders (Ghosh et al., 2020; Nahum-Shani et al., 2020a), as in the current design. This restriction is typically informed by prior empirical evidence or practical considerations indicating that individuals who show early signs of response are likely to benefit from continuing with the initial intervention, whereas those showing early signs of non-response are unlikely to benefit and may require treatment modification in order to succeed (Nahum-Shani & Almirall, 2019). The randomization of non-responders at week 2 in the current example is designed to investigate whether this subgroup would benefit from adding meal replacement (Meal).

Suppose that 400 individuals enter the study (for simplicity, throughout we assume no attrition) and are randomized at program entry with equal probability (0.25) to each of the four experimental options 1-4 in Figure 1a (i.e., $n$=100 participants in each option). Then, suppose that at week 2, 50% of the participants are classified as responders ($n$=200) and continue with the initial intervention (option A). Those classified as non-responders ($n$=200) are re-randomized with equal probability (0.50) to either add meal replacement (option B; $n$=100) or not (option C; $n$=100).

The sequential randomization in Figure 1a leads to 12 experimental cells, which are detailed in Figure 1b. For example, participants who start with both Coaching and App (option 1 at program entry) and then show early signs of response and thus continue with the initial intervention (option A at week 2) are considered part of experimental cell 1→A; those who started with both Coaching and App (option 1 at program entry) and then show early signs of non-response and are assigned to add Meal (option B at week 2) are considered part of experimental cell 1→B. As we discuss below, similar to a standard factorial design, the goal here is not to compare these individual cells, but rather to combine multiple cells in different ways to answer multiple scientific questions about the selection of intervention components (Collins, 2018).

Further, similar to a prototypical SMART design (Nahum-Shani et al., 2020a), several adaptive interventions are embedded in this trial. An embedded adaptive intervention is a sequence of decision rules that is included in the trial by design. It specifies which components to deliver (and for whom) at each decision point (see Nahum-Shani & Almirall, 2019; Inbal Nahum-Shani et al., 2022). Here, there are eight embedded adaptive interventions, which are described in Table 1; each is represented by 2 of the 12 experimental cells in Figure 1b, with some adaptive interventions sharing cells. Next, we describe how data from this hypothetical study can be analyzed to answer questions Q1-Q3, as well as other, more exploratory questions.

**Data analytic approach for hybrid factorial-SMART.** In the current example the primary outcome of interest is weight loss measured at the month 6 follow up. Data from an illustrative (hypothetical) experimental design shown in Figure 1 can be used to answer the three motivating questions outlined above, about the inclusion of App and Coaching at the beginning of the program and the inclusion of Meal for non-responders at week 2. This can be done by testing the main effect of each corresponding factor. Specifically, using data from the HED in Figure 1, the main effect of *App* can be estimated by comparing the mean outcome across all the cells in which *App* was set to On (the 6 cells starting with options 1 and 2; $n=200$; Figure 1) to the mean outcome across all the cells in which *App* was set to Off (the 6 cells starting with options 3 or 4; $n=200$; Figure 1). Similarly, the main effect of *Coaching* can be estimated by comparing the 6 cells starting with options 1 and 3 ($n=200$) to the 6 cells starting with options 2 and 4 ($n=200$). Notice that both main effects are estimated by using outcome information from the entire sample ($N=400$), and they average over the factor *Meal* and the response status.

Next, the main effect of adding meal replacement to non-responders (*Meal*) can be estimated in two ways: the first is conditional on response status, and the second is marginal (i.e., averages over response status). The conditional approach would be restricted to non-responders and would involve comparing the mean outcome of all non-responders who were offered Meal at week 2 (i.e., the 4 cells involving option B in Figure 1b; $n$=100) with the mean outcome of all non-responders who were not offered Meal at week 2 (i.e., the 4 cells involving option C in Figure 1b; $n$=100). Notice that this main effect is estimated by using outcome information from the entire sample of non-responders ($n$=200), but it only includes non-responders. Thus, this main effect is averaged over the factors assigned at program entry, but it is conditional on response status. Alternatively, the marginal approach would involve comparing (a) the mean outcome across all the cells of responders (i.e., the 4 cells involving option A in Figure 1b; $n$=200) and non-responders who were offered Meal at week 2 (i.e., the 4 cells involving option B in Figure 1b; $n$=100) with (b) the mean outcome across all the cells of responders (i.e., the 4 cells involving option A in Figure 1b; $n$=200) and non-responders who were not offered Meal at week 2 (i.e., the 4 cells involving option C in Figure 1b; $n$=100). Notice that this main effect is estimated by using outcome information from the entire sample, including responders and non-responders. Thus, this definition of the main effect is averaging over the factors assigned at program entry *and* response status. The conditional effect will generally be a larger quantity than the marginal effect, because *Meal* has no effect on responders, and therefore the marginal effect is a weighted average of the (potentially nonzero) effect on non-responders and the (necessarily zero) effect on responders. However, as we discuss below, this does not matter for testing purposes. Below, we discuss modeling and estimation considerations in using data from a hybrid factorial-SMART to estimate main effects, as well as interactions between factors.

***Proposed model and estimands.*** Let $Z_{11}$ and $Z_{12}$ be the randomized factors at program entry (here, *App* and *Coaching*, respectively), both coded $+1$ when the corresponding component is set to On and $-1$ when it is set to Off, and let $Z_{21}$ be the randomized factor for non-responders at week 2 (here, *Meal*), also coded $+1$ for On and $-1$ for Off. Let $\boldsymbol{X}_0$ be a vector of control covariates measured prior to first-stage randomization (and included in the regression models as mean-centered covariates). The following is a relatively simple model that can be fitted to answer questions about the main effect of each component and interactions between them:

$$E(Y|\boldsymbol{X}_0, Z_{11}, Z_{12}, Z_{21}) = \boldsymbol{X}_0\boldsymbol{\gamma}_0 + \theta_1 Z_{11} + \theta_2 Z_{12} + \theta_3 Z_{11}Z_{12} + \tag{1}$$
$$\theta_4 Z_{21} + \theta_5 Z_{11}Z_{21} + \theta_6 Z_{12}Z_{21} + \theta_7 Z_{11}Z_{12}Z_{21}.$$

Throughout we assume for simplicity that the outcome is continuous, but all the models discussed in this manuscript can be extended to other types of outcomes (e.g., binary and count) by using the appropriate link function.

Table 2 summarizes how the coefficients in Model (1) can be used to answer a variety of scientific questions about the main effect of each component and the interactions between them. For example, consider Q1 above, which concerns whether at program entry it is beneficial (in terms of weight loss at month 6), to offer the App (vs. no App). Based on Model (1), $E(Y|Z_{11} = 1) = \theta_1$ is the average weight loss for those who were offered App at program entry, averaging over response status and the other factors (i.e., $Z_{12}$ and $Z_{21}$). Similarly, $E(Y|Z_{11} = -1) = -\theta_1$ is the average weight loss for those who were not offered App, and so $\left(E(Y|Z_{11} = 1) - E(Y|Z_{11} = -1)\right) = 2\theta_1$ is the main effect of *App*. This main effect is the average difference in weight loss between those who were offered the app and those who were not offered the app at program entry, averaging over response status and the other factors. Similarly, $2\theta_2$ can be used to estimate the main effect of *Coaching*, which is the focus of Q2.

Further, $2\theta_4$ can be used to answer Q3, which concerns whether it would be beneficial to add (vs. not add) meal replacement for individuals who show early signs of non-response at week 2; this quantity represents the main effect of *Meal*, averaging over the other factors and response status. This main effect is based on the marginal approach discussed above, which averages over response status. Estimating the effect of *Meal* only among non-responders (i.e., conditional on response status) would require rescaling the quantity $2\theta_4$ by a factor of $1/(1 - r)$, where $r$ is the response rate. The magnitude of this conditional effect will generally be larger than the marginal effect, but the statistical tests are equivalent because both the estimand and standard error would be rescaled by the same factor, leading to the same conclusions (e.g., same $p$-value).

The coefficients in Model (1) can also be used to answer scientific questions about how well the components work together. For example, is the app more beneficial with or without coaching? The interaction between the two initial factors $Z_{11}$ and $Z_{12}$ can be used to answer this question. Specifically, this interaction represents the difference between two "simple effects": (1) the effect of *App* when *Coaching* is set to On, averaging over response status and $Z_{21}$, namely $\left(E(Y|Z_{11} = 1, Z_{12} = 1) - E(Y|Z_{11} = -1, Z_{12} = 1)\right) = 2\theta_3$; and (2) the effect of *App* when *Coaching* is set to Off, averaging over response status and $Z_{21}$, namely $\left(E(Y|Z_{11} = 1, Z_{12} = -1) - E(Y|Z_{11} = -1, Z_{12} = -1)\right) = -2\theta_3$. The difference between these two simple effects is $4\theta_3$; rescaling this quantity by ½ for comparability (see Collins et al., 2018), $2\theta_3$ represents the interaction between *App* and *Coaching*.

Next, consider another scientific question about how well components that are offered at different decision points work together: Does the overall benefit of the App (in terms of month 6 weight loss) change based on whether Meal is offered to non-responders at week 2? The marginal interaction between $Z_{11}$ and $Z_{21}$ can be used to answer this question. This interaction represents the difference between two "simple effects": (1) the effect of *App* when *Meal* is set to On among non-responders, averaging over response status and $Z_{12}$, namely $\left(E(Y|Z_{11} = 1, Z_{21} = 1) - E(Y|Z_{11} = -1, Z_{21} = 1)\right) = 2\theta_5$; and (2) the corresponding effect when *Meal* is set to Off among non-responders, namely $-2\theta_5$. The difference between these two simple effects is $4\theta_5$; rescaling this quantity by ½ for comparability (see Collins et al., 2018), $2\theta_5$ represents the interaction between *App* and *Meal*, averaging over response status and *Coaching*.

Finally, linear combinations of the coefficients in Model 1 can also be used to compare adaptive interventions that are embedded in the design. The contrast between expected population outcomes under embedded adaptive interventions $(z_{11}, z_{12}, z_{21})$ and $(z'_{11}, z'_{12}, z'_{21})$ is

$$
\begin{aligned}
&(\theta_1 z_{11} + \theta_2 z_{12} + \theta_3 z_{11} z_{12} + \theta_4 z_{21} + \theta_5 z_{11} z_{21} + \theta_6 z_{12} z_{21} + \theta_7 z_{11} z_{12} z_{21}) \\
-&(\theta_1 z'_{11} + \theta_2 z'_{12} + \theta_3 z'_{11} z'_{12} + \theta_4 z'_{21} + \theta_5 z'_{11} z'_{21} + \theta_6 z'_{12} z'_{21} + \theta_7 z'_{11} z'_{12} z'_{21}) \\
=& \ \theta_1(z_{11} - z'_{11}) + \theta_2(z_{12} - z'_{12}) + \theta_3(z_{11} z_{12} - z'_{11} z'_{12}) + \theta_4(z_{21} - z'_{21}) + \\
& \theta_5(z_{11} z_{21} - z'_{11} z'_{21}) + \theta_6(z_{12} z_{21} - z'_{12} z'_{21}) + \theta_7(z_{11} z_{12} z_{21} - z'_{11} z'_{12} z'_{21}).
\end{aligned}
$$

For example, consider the contrast between $(+1, 1, 1)$ and $(-1, 1, 1)$, which can be quantified based on Model 1 as follows: $\mu(+1, 1, 1) - \mu(-1, 1, 1) = 2\theta_1 + 2\theta_3 + 2\theta_5 + 2\theta_7$. This is a contrast between two embedded adaptive interventions, both offering Coaching initially and then offering Meal replacement to non-responders (responders continue with the initial intervention), but the former also offers App initially (adaptive intervention #1 in Table 1), whereas the later does not (adaptive intervention #5 in Table 1).

***Estimation.*** The Weight and Replicate (W&R) data analysis method (Dziak et al., 2019; Nahum-Shani et al., 2020a; Nahum-Shani et al., 2012a) can be used to estimate the parameters in Model (1). Appendix A of Nahum-Shani et al., (2012a) provides a technical explanation of this approach and demonstrates that the estimators are unbiased in large samples. W&R was developed to address two challenges related to estimation in a prototypical SMART setting; these challenges are also relevant to the hybrid factorial-SMART in Figure 1. First, responders are not re-randomized to the second-stage factor (*Meal*); instead they are offered a fixed intervention option (continue with the initial intervention). Thus, the value of $Z_{21}$ for responders is not specified (by design). Further, each responder provides data that are consistent with two of the eight embedded adaptive interventions in Table 1. For example, responders to App alone (i.e., those in Cell 2→A) provide data that are consistent with both embedded adaptive intervention #3 (Table 1) and embedded adaptive intervention #4 (Table 1). Thus, the value of $Z_{21}$ should be specified in such a way that enables outcome data from responders to be used to estimate the mean outcome under both embedded adaptive interventions. W&R employs *replication* of responders' data to address this challenge.

Specifically, each row in the dataset that pertains to a responder is duplicated, using the same values for all variables (including $Y$), except for $Z_{21}$. One of the duplicated rows is assigned $Z_{21} = 1$, while the other is assigned $Z_{21} = -1$. The number of rows in the new, re-structured dataset will be $N + \sum_{i=1}^{N} R_i$, where $N$ is the actual number of original participants before replication, and $R$ indicates whether $(R = 1)$ or not $(R = 0)$ the $i'$th participant was classified as a responder to first-stage components. It is important to note that W&R does not assume that the true data set is of size $N + \sum_{i=1}^{N} R_i$; rather, the method correctly assumes that the sample size is $N$ for purposes of calculating standard errors (see Nahum-Shani et al., 2020a; Nahum-Shani et al., 2012a).

Second, non-responders are re-randomized to second-stage intervention options, while responders are not re-randomized (in Figure 1 they continue with the option assigned at program entry). Therefore, outcome information from non-responders is underrepresented in the sample mean under a particular embedded adaptive intervention; this underrepresentation occurs by design. Correspondingly, the probability of being assigned to a sequence of components that is consistent with a particular embedded adaptive intervention is lower for non-responders than responders. In Figure 1 this probability is 0.25 for non-responders (because they were randomized twice with probability 0.5 each time) and 0.5 for responders (because they were randomized only once with probability 0.5). Because of this imbalance, and because response status is expected to be associated with the outcome, taking a naïve average of $Y$ for all individuals consistent with a specific embedded adaptive intervention likely leads to bias (Orellana et al., 2010; Robins et al., 2008). To restore balance, W&R employs weights that are proportional to the inverse of the probability of treatment assignment. In the case of the hybrid factorial-SMART in Figure 1, responders' observations are assigned the weight $W$=2 (the inverse of 0.5), and non-responders' observations are assigned the weight $W$=4 (the inverse of 0.25). These weights are often referred to as "known weights" because they are based on the design's known treatment assignment probabilities. Alternatively, these weights can be estimated based on data from the SMART, with the potential to improve the estimator's efficiency (i.e., achieve smaller standard errors; Brumback, 2009; Hernan et al., 2002; Hirano et al., 2003; Nahum-Shani et al., 2020a).

With the data replicated and weighted as described above, Model (1) can be estimated by using standard software (e.g., SAS, R) with robust (sandwich) standard errors (e.g., in SAS GENMOD procedures). Nahum-Shani et al (2012a) provide a detailed justification for using robust standard errors in this setting.

## Hybrid Factorial-MRT

Suppose investigators are motivated to answer the following questions about the development of a weight loss intervention that offers a mobile app to participants at the beginning of the program. The first two questions concern whether or not, at the beginning of the program, the app should be augmented with other components. Specifically, at the beginning of the program is it beneficial (in terms of weight loss at month 6), to (Q4) augment the app with coaching sessions; and/or (Q5) augment the app with meal replacement. The third question concerns the delivery of a prompt to encourage monitoring of that day's dietary intake. Specifically, (Q6) is it beneficial (in terms of increasing self-monitoring of dietary intake in the next 12 hours) to deliver a mobile-based prompt (via a push notification) at mid-day? Note that Q4 and Q5 focus on components that can be delivered at a single decision point (here, at program entry). On the other hand, Q6 concerns a time-varying component—a component that may or may not be delivered at several decision points during the course of the intervention (i.e., the decision whether or not to deliver this component is made repeatedly during the intervention).

To answer these questions, the investigators may consider an experiment with three factors, one factor for each component. Here, *Add Coaching* refers to the factor corresponding to augmenting the app with coaching throughout the study, *Add Meal* refers to the factor corresponding to augmenting the app with meal replacement throughout the study, and *Prompt* refers to the factor corresponding to delivering a prompt to motivate self-monitoring on a given day. Each factor will have two levels: On and Off. Since decisions regarding the components Add Coaching and Add Meal should be made at the beginning of the program, study participants should be randomized to the corresponding factors *Add Coaching* and *Add Meal* at a single decision point – just prior to beginning the program. However, decisions regarding the delivery of the component Prompt can be made repeatedly, every day. Hence, study participants should be randomized to the corresponding factor *Prompt* at multiple decision points (i.e., daily at mid-day). We use the term "time-varying factor" to refer to a factor to which participants can be randomly assigned at multiple decision points during the study.

Note that Q4 and Q5 (about the benefits of augmenting the app with coaching and meal replacement at program entry) focus on a distal outcome—weight loss at month 6, whereas Q6 (about the benefits of delivering a mobile-based prompt at mid-day) focuses on a proximal outcome—self-monitoring of dietary intake in the next 12 hours. The focus of Q6 on a proximal outcome is motivated by the timescale at which the component of interest can be delivered (every day) and the timescale at which it is likely to impact behavior change (the delivery of a mobile-based prompt on a given day is likely to encourage relatively near-term self-monitoring behaviors).

The experimental design in Figure 2 can provide data for addressing questions Q4-Q6. At the beginning of the study, participants are randomized to one of four experimental options resulting from crossing the levels of the two factors *Add Coaching* and *Add Meal*. In addition, throughout the first 12 weeks, individuals are randomized daily to the two levels of the factor *Prompt*. This design integrates a standard 2×2 factorial experiment with a micro-randomized trial (MRT; Liao et al., 2016; Qian et al., 2022). An MRT is a form of a factorial design (Walton et al., 2018) involving *rapid* sequential randomization to experimental factors. Specifically, randomizations to experimental factors occur frequently, and the time interval between randomizations is relatively short (e.g., hourly, daily). An MRT with a single factor that can take either of two levels (e.g., prompt vs. no prompt) at each randomization time can be viewed as a factorial design with $2^K$ potential experimental cells, where $K$ is the number of decision points in which randomizations take place. However, similar to a standard factorial design, the primary scientific focus is *not* on comparing individual cells, but rather on the average main effect of the factor on a proximal outcome (i.e., proximal main effect), which, as we explain below, can be estimated by pooling outcome data across all $K$ decision points and $N$ individuals.

Suppose that 400 individuals enter the study and are randomized initially with equal probability (0.25) to each of the four conditions 1-4 in Figure 2 (i.e., $n$=100 participants in each condition). Then, suppose that every day, at mid-day, each individual is randomized with equal probability (0.5) to each of the two levels of *Prompt*, that is, to either a prompt (condition A; $n$=100) or no prompt (condition B; $n$=100). Similar to a standard factorial design, the goal here is not to compare the cells resulting from crossing the levels of all factors at all decision points that involve randomization, but rather to combine multiple cells in different ways to answer multiple scientific questions about the selection of intervention components (Collins, 2018). Below, we discuss how data from this hybrid Factorial-MRT design can be used to answer questions Q4-Q6.

**Data analytic approach for hybrid factorial-MRT.** Recall that in the current example, questions Q4 and Q5 (about the benefits of adding coaching to the app and adding meal replacement to the app) concern a distal outcome—weight loss measured at the month 6 follow up. Data from the experimental design in Figure 2 can be used to answer these questions by testing the distal main effect of each corresponding factor. Recall that when a factor has two levels, the main effect of this factor can be defined as the difference between the mean outcome at one level of this factor and the mean outcome at the other level of this factor, averaging over the levels of the other factors. Thus, using data from the HED in Figure 2, the main effect of *Add Coaching* can be estimated by comparing the mean outcome across all the cells in which *Add Coaching* was set to On (i.e., those starting with options 1 and 2 in Figure 2; $n$=200) to the mean outcome across all the cells in which *Add Coaching* was set to Off (i.e., those starting with options 3 or 4 in Figure 2; $n$=200). Similarly, the main effect of *Add Meal* can be estimated by comparing the mean outcome across all the cells in which *Add Meal* was set to On (i.e., those starting with options 1 and 3 in Figure 2; $n$=200) to the mean outcome across the cells in which *Add Meal* was set to Off (i.e., those starting with options 2 and 4 in Figure 2; $n$=200). Note that both main effects also average over the levels of the factor *Prompt* at each decision point (1 through 84), and both are estimated by using outcome information from the entire sample ($N$=400).

Next, recall that Q6, about the benefits of delivering a prompt to encourage daily self-monitoring, concerns a proximal outcome—self-monitoring of dietary intake over the next 12 hours. This proximal outcome is assessed following each randomization. The main effect of the time-varying factor *Prompt* can be estimated by comparing the mean outcome across all days in which a prompt was delivered (i.e., option A which was initiated, on average, on half of the days) to the mean outcome across all days in which a prompt was not delivered (i.e., option B which was initiated, on average, on half of the days). This difference is estimated by pooling data not only across all decision points in the trial, but also across all study participants (Qian et al., 2022). That is, proximal outcome data from participants in all four initial experimental options 1-4 ($N$=400) can be used to estimate this effect, which represents the (causal) main effect of *Prompt* in terms of the proximal outcome, averaging over the other factors and all decision points in the trial.

Notably, in a standard factorial design, the focus is on estimating the distal main effect of a factor, namely the main effect of a given factor on an outcome that is typically measured at a single (or a few) fixed later end-points (e.g., an end-of-study follow up), averaging over the levels of the other factors in the design (Collins et al., 2009). However, in standard MRTs the main focus is on estimating the average proximal main effect of a factor, namely the main effect of a given factor on an outcome that is measured rapidly, following each decision point, averaging over all decision points (and individuals) in the trial (Qian et al., 2022). By integrating features from a standard factorial design and a standard MRT, a hybrid factorial-MRT provides data that can be used to estimate both distal main effects and proximal main effects, averaging over the levels of the other factors in the experiment, as well as over the decision points in the experiment. Below, we discuss modeling and estimation considerations in using data from a hybrid factorial-MRT to estimate main effects as well as interactions between factors.

***Proposed model and estimands.*** Let $Z_{11}$ and $Z_{12}$ be the two factors randomized at program entry (here, *Add Coaching* and *Add Meal*, respectively), both coded $+1$ for On and $-1$ for Off. Let $t = 1, \dots, T$ be the decision points at which individuals are micro-randomized; in the current example, there is one decision point per day over 12 weeks: $T = 7 \times 12 = 84$ days. Let $A_{it}$ represent the micro-randomized factor (here, *Prompt*) for individual $i$, at time $t$, with levels $+1$ for On (i.e., prompt) and $-1$ for Off (i.e., no prompt). Let $Y_i^*$ represent the distal outcome (here, weight loss at month 6) for individual $i$ and let $Y_{it+\Delta}$ represent the proximal outcome (self-monitoring dietary intake) for individual $i$ at time $t + \Delta$. In the current example, $\Delta$ represents the next 12 hours following decision point $t$; that is, $\Delta$ does not represent a separate model parameter, but rather some unspecified time $\Delta$ that comes after the assignment of $A_{it}$ but before the assignment of $A_{it+1}$. Thus, $A_{it}$ may causally affect $Y_{it+\Delta}$. As before, let $\boldsymbol{X}_0$ be a vector of control covariates measured prior to program entry randomizations and included in the regression models as mean-centered covariates. The following model can be estimated to answer questions about the distal main effect (Q4 and Q5 outlined above) of the factors assigned at program entry and the interactions between them:

$$E(Y_i^* | \boldsymbol{X}_{i0}, Z_{i11}, Z_{i12}) = \boldsymbol{X}_{i0}\boldsymbol{\theta}_0 + \theta_1 Z_{i11} + \theta_2 Z_{i12} + \theta_3 Z_{i11}Z_{i12}.$$

This model is marginal over $\bar{A}_i$, which is the average of the factor *Prompt* (i.e., the rate of prompt delivery) across all decision points for individual $i$. Alternatively, the subject-specific average of the factor *Prompt* can be included in the model as follows:

$$\begin{aligned} E(Y_i^* | \boldsymbol{X}_{i0}, Z_{i11}, Z_{i12}, \bar{A}_i) = \ & \boldsymbol{X}_{i0}\boldsymbol{\theta}_0 + \theta_1 Z_{i11} + \theta_2 Z_{i12} + \theta_3 Z_{i11}Z_{i12} \\ & + \theta_4 \bar{A}_i + \theta_5 Z_{i11}\bar{A}_i + \theta_6 Z_{i12}\bar{A}_i + \theta_7 Z_{i11}Z_{i12}\bar{A}_i. \end{aligned} \tag{2}$$

Table 3 summarizes how the coefficients in Model (2) can be used to answer a variety of scientific questions about the distal main effects of the factors assigned at program entry, the interactions between them, and their interactions with the time-varying factor. For example, consider Q4: whether at the beginning of the program it is beneficial (in terms of weight loss at month 6) to add (vs. not add) coaching. Assuming participants are assigned to each level of $Z_{11}$, $Z_{12}$ and $A$ with equal probability (0.5), then $E(Z_{11}) = 0$, $E(Z_{12}) = 0$, $E(Z_{11}Z_{12}) = 0$, and $E(\bar{A}) = 0$. Therefore, based on Model (2), $E\big(E(Y^*|Z_{11} = 1) - E(Y^*|Z_{11} = -1)\big) = 2\theta_1$ is the main effect of *Add Coaching*, namely the average difference in weight loss between those who were offered coaching and those who were not offered coaching at the beginning of the program, averaging over the other factor that was assigned at program entry (*Add Meal*) and the time-varying factor (*Prompt*). Similarly, $E\big(E(Y^*|Z_{12} = 1) - E(Y^*|Z_{12} = -1)\big) = 2\theta_2$ can be used to answer Q5, which concerns whether at the beginning of the program it is beneficial to add (vs. not add) meal replacement. This quantity represents the main effect of *Add Meal*, averaging over the other factor that was assigned at program entry (*Add Coaching*) and the time-varying factor (i.e., *Prompt*).

The parameters in Model (2) can be used to answer other scientific questions about the interactions between the factors assigned at program entry and the time-varying factor in terms of the distal outcome. For example, consider Question D from Table 3: Does the effect of (i.e., difference between) adding (vs. not adding) coaching at the beginning of the program on the distal outcome (weight loss by month 6) vary by the rate of prompt delivery? This question concerns the *interaction* between a factor that was assigned at program entry and a time-varying factor that was assigned daily, in relation to the distal outcome. The randomizations to the time-varying factor *Prompt* (prompt vs. no prompt) with ½ probability each day lead to slightly different total numbers of prompts delivered per participant. Although the values of this distribution may be tightly clustered around the mean (due to the law of large numbers), they do differ slightly by random chance from participant to participant. Based on Model 2, subtracting the conditional effects of the factor *Add Coaching* on the distal outcome, given two different $\bar{A}$ values of interest, can be interpreted as an interaction. For example, if $\bar{A} = 0.4$, then the conditional effect of *Add Coaching* is $\big(E(Y^*|Z_{11} = +1, \bar{A} = .4) - E(Y^*|Z_{11} = -1, \bar{A} = .4)\big) = (\theta_1 + 0.4\theta_5) - (-\theta_1 - 0.4\theta_5) =$ $2(\theta_1 + 0.4\theta_5)$, and if $\bar{A} = 0.1$, then the conditional effect is $2(\theta_1 + 0.1\theta_5)$. The difference between these two conditional effects is $2 \times 0.3\theta_5$. Rescaling this quantity by ½ (Collins et al., 2018), $0.3\theta_5$ represents the interaction between *Add Coaching* and the specified difference in rate of prompt delivery. As before, this interaction would be estimated by using data across all study participants and across all decision points.

Next, the following model can be estimated to answer questions about the proximal main effect of the time-varying factor (*Prompt*) and the interactions between this factor and the two factors assigned at program entry (*Add Coaching* and *Add Meal*).

$$E(Y_{it+\Delta}|\boldsymbol{X}_{i0}, Z_{i11}, Z_{i12} A_{it}) = \boldsymbol{X}_{i0}\boldsymbol{\beta}_0 + \beta_1 Z_{11} + \beta_2 Z_{i12} + \beta_3 Z_{i11}Z_{i12} \\ + \gamma_0 A_{it} + \gamma_1 Z_{i11}A_{it} + \gamma_2 Z_{i12}A_{it} + \gamma_3 Z_{i11}Z_{i12}A_{it} \tag{3}$$

Table 4 summarizes how the coefficients in Model (3) can be used to answer a variety of scientific questions about the proximal main effect of the time-varying component, as well as other questions about the interaction between this component and the two program entry components. For example, consider Q6, which concerns whether it is beneficial (in terms of self-monitoring of dietary intake in the next 12 hours) to deliver a prompt at mid-day. This question concerns the main effect of the time-varying factor *Prompt* on the proximal outcome, averaging over the other factors and all decision points in the trial. Under Model (3), $E(Y_{it+\Delta}|A_{it} = +1) - E(Y_{it+\Delta}|A_{it} = -1) = 2\gamma_0$ is the proximal main effect of *Prompt,* averaging over the other factors and decision points in the trial.

The parameters in Model (3) can also be used to answer scientific questions about the interaction between the time-varying factor and other factors in terms of the proximal outcome. For example, consider Question B from Table 4: Does the effect of *Prompt* (i.e., the difference between delivering vs. not delivering a prompt) on the proximal outcome vary by whether or not coaching was initiated at program entry? This question concerns the *interaction* between the time-varying factor and a factor that was assigned at program entry in relation to the proximal outcome. Using Model 3,

$$
\begin{aligned}
\big(E(Y_{it+\Delta}|A_{it} = +1, Z_{i1}\ \ = +1) - E(Y_{it+\Delta}|A_{it} = -1, Z_{i11} = +1)\big)& \\
-\big(E(Y_{it+\Delta}|A_{it} = +1, Z_{i11} = -1) - E(Y_{it+\Delta}|A_{it} = -1, Z_{i11} = -1)\big)& = 4\gamma_1.
\end{aligned}
$$

Rescaling this quantity by ½ (Collins et al., 2018), $2\gamma_1$ represents the interaction between *Prompt* and *Add Coaching*, averaging over *Add Meal* and all the decision points in the trial. Other scientific questions and their associated Model (3) parameters are presented in Table 4.

   ***Estimation.*** Model (2) can be estimated using standard regression procedures that are included in standard statistical software (e.g., SAS, R). Model (3) can be estimated with repeated measurement regression procedures in standard statistical software, such as Generalized Estimating Equations (GEE; Liang & Zeger, 1986). However, if the model includes time-varying covariates that depend on previous outcomes or previous interventions (e.g., prior self-monitoring behaviors or craving measured prior to decision point $t$), these methods can result in inconsistent estimators (i.e., estimators that are different from the true parameter value even in large samples) for the proximal effects of $A_{it}$ (see Qian et al., 2022; Qian et al., 2020). In this case, the weighted and centered least-squares (WCLS) estimation procedure developed by Boruvka and colleagues (2018) can be employed. Note that while WCLS provides consistent estimators of the proximal effects of $A_{it}$ (marginal or conditional on $Z_{i11}$ and $Z_{i12}$), it may not ameliorate the problem for other proximal effects (e.g., the proximal main effect of $Z_{i11}$). An accessible overview of this method is provided by Qian and colleagues (2022).

## Hybrid SMART-MRT

   Suppose investigators are motivated to answer the following questions about the development of a weight loss intervention that offers a mobile app to participants at the beginning of the program. First (Q7), at the beginning of the program is it beneficial (in terms of weight loss at month 6) to augment the app with coaching (Add Coaching)? Next (Q8), at week 2, is it beneficial (in terms of weight loss at month 6) to add meal replacement (Add Meal) to individuals who show early signs of non-response? Finally (Q9), is it beneficial (in terms of increasing self-monitoring of dietary intake in the next 12 hours) to deliver a mobile-based prompt at mid-day (Prompt)? Note that Q7 and Q8 concern different components that can each be initiated at a single, unique decision point—at program entry (Add Coaching) and at week 2 (Add Meal), respectively. On the other hand, Q9 concerns a time-varying component, and the decision to deliver it or not can be made daily (Prompt). Also notice that, consistent with the timescale at which the component of interest can be delivered, Q7 and Q8 focus on a distal outcome (weight loss at month 6) whereas Q9 concerns a proximal outcome (self-monitoring of dietary intake in the next 12 hours).

To answer these questions, investigators may consider an experiment with three factors, one for each component. Here, *Add Coaching* refers to the factor corresponding to augmenting the app with coaching at program entry, *Add Meal* refers to the factor corresponding to adding meal replacement to non-responders at week 2, and *Prompt* refers to the factor corresponding to delivering a mid-day prompt to motivate self-monitoring. Each factor will have two levels: On and Off. Since decisions regarding the component Add Coaching should be made at the beginning of the program, study participants should be randomized to the corresponding factor *Add Coaching* at a single decision point (the beginning of the program). Further, since decisions regarding the component Add Meal should be made at week 2 and they only concern non-responders, the randomization to the corresponding factor *Add Meal* should take place at a single decision point (week 2) and involve only non-responders. Finally, since decisions about the delivery of the component Prompt can be made repeatedly, participants should be randomized to the corresponding factor *Prompt* at multiple decision points (i.e., daily at mid-day).

The experimental design in Figure 3a can provide data for addressing questions Q7-Q9. At program entry, participants are randomized to the two levels of the factor *Add Coaching*. Then, at week 2, non-responders are randomized to the two levels of the factor *Add Meal*, whereas responders continue with their assigned initial option. In addition, throughout the first 12 weeks, individuals are randomized daily, at midday, to the two levels of the factor *Prompt*. This design integrates a prototypical SMART with an MRT (Nahum-Shani et al., 2022).

Suppose that 400 individuals enter the study and are randomized initially with equal probability (0.50) to the two levels of *Add Coaching* (i.e., $n$=200 participants in each option). Then, suppose that at week 2, 50% of the participants (i.e., $n$=200) are classified as responders ($n$=200 in option A from Figure 3a) and continue with the initial intervention. Those classified as non-responders ($n$=200) are re-randomized with equal probability (0.50) to either add Meal ($n$=100 in option B from Figure 3a) or not ($n$=100 in option C from Figure 3a). Then, suppose that every day, at mid-day, each individual is randomized with equal probability (0.5) to each of the two levels of *Prompt*, that is to either deliver a prompt ($n$=100) or not ($n$=100).

Note that the sequential randomization in Figure 3a leads to 6 experimental cells, which are detailed in Figure 3b. For example, participants who start with both App alone (option 1 at program entry) and then are classified as responders at week 2 and thus continue with the initial intervention (option A at week 2) are considered part of experimental cell 1→A. However, as before, the goal here is not to compare these cells but rather to combine multiple cells in different ways to answer the scientific questions outlined above (Collins, 2018). Further, similar to a prototypical SMART design (Nahum-Shani et al., 2020a), there are four adaptive interventions that are embedded in this trial by design (see Table 5). However, it is important to note that in a hybrid SMART-MRT the adaptive interventions are embedded in the context of a pre-specified protocol for delivering a time-varying component (here the delivery of prompts with 0.5 probability each day) and thus their effects should be interpreted accordingly. Below, we discuss how data from this hybrid SMART-MRT design can be used to answer questions Q7-Q9.

**Data analytic approach for hybrid SMART-MRT.** A data analysis approach for the hybrid SMART-MRT would combine the special features of the hybrid factorial-SMART and hybrid factorial-MRT designs. Like the factorial-SMART design, the distinction between responders and non-responders is important, particularly when (as in our example) the re-randomization in the SMART is restricted to non-responders. Further, similar to the factorial-MRT design, the distinction between proximal and distal outcomes is also important. The notation, however, does not become much more complicated, and the underlying ideas are very similar.

***Proposed model and estimands.*** Let $Z_{i1}$ be the randomized factor at program entry (here, *Add Coaching*) and $Z_{i2}$ be the randomized factor at week 2 for non-responders (here, *Add Meal*), each coded $+1$ for On and $-1$ for Off. As before, let $\boldsymbol{X_0}$ denote mean-centered baseline covariates. Finally, let $A_{it}$ represent the micro-randomized factor (here, *Prompt*) for individual $i$ at time $t$, coded $+1$ for On (i.e., prompt) and $-1$ for Off (i.e., no prompt). As in the hybrid factorial-MRT, let $Y_{it\ \Delta}$ represent the proximal outcome (here, self-monitoring in the next 12 hours) and $Y_i^*$ represent the distal outcome (here, month 6 weight loss). The following model can be estimated to answer questions about the distal main effects of the factor assigned at program entry and the factor assigned at week 2, as well as the interaction between them.

$$E(Y_i^* | \boldsymbol{X}_{i0}, Z_{i1}, Z_{i2}) = \boldsymbol{X}_{i0}\boldsymbol{\theta}_0 + \theta_1 Z_{i1} + \theta_2 Z_{i2} + \theta_3 Z_{i1} Z_{i2} \tag{4}$$

Model 4 is marginal over $\bar{A}_i$ and it has essentially the same form as the marginal model for the distal outcome in a SMART without an MRT component (see Nahum-Shani et al., 2012a). As in the case of a hybrid factorial-MRT, a more general model could instead be used to include an effect of $\bar{A}_i$, with a similar caveat that due to the law of large numbers, there will be little between-subject variability in $\bar{A}_i$ (see Model 2 in Appendix A of Nahum-Shani et al., 2022).

Table 6 summarizes how the coefficients in Model (4) can be used to answer questions about the distal effects of the $Z_{i1}$ and $Z_{i2}$. For example, the expected distal outcome for a participant who was offered coaching at program entry is $E(Y_i^* \mid Z_{i1} = +1) = \theta_1$, and the expected distal outcome for a participant who was not offered coaching at program entry is $E(Y_i^* \mid Z_{i1} = -1) = -\theta_1$. Therefore, the difference between these quantities, namely $2\theta_1$ can be used to answer question Q7, which concerns the main effect of the factor *Add Coaching*. Likewise, $2\theta_2$ is the main effect of the factor *Add Meal* for non-responders, which is the focus of question Q8. Finally, as in Model 1, linear combinations of the coefficients in Model 4 can be used to compare adaptive interventions that are embedded in the hybrid SMART-MRT, averaging over the time-varying factor. Specifically, the contrast between expected population outcomes under embedded adaptive interventions $(z_1, z_2)$ and $(z_1', z_2')$ is

$$(\theta_1 z_1 + \theta_2 z_2 + \theta_3 z_1 z_2) - (\theta_1 z_1' + \theta_2 z_2' + \theta_3 z_1' z_2')$$
$$= \theta_1(z_1 - z_1') + \theta_2(z_2 - z_2') + \theta_3(z_1 z_2 - z_1' z_2').$$

Having considered the distal effects of the factors $Z_{i1}$ and $Z_{i2}$, we now consider the proximal effects of the time-varying factor. The simplest model for answering question Q9, which concerns the effect of *Prompt* on self-monitoring in the next 12 hours, would be

$$E(Y_{it+\Delta} \mid \boldsymbol{X}_{i0}, Z_{i1}, Z_{i2}, A_{it}) = \boldsymbol{X}_{i0}\boldsymbol{\beta}_0 + \beta_1 Z_{i1} + \beta_2 C_{it} Z_{i2} + \beta_3 C_{it} Z_{i1} Z_{i2} \qquad (5)$$
$$+ \gamma_0 A_{it} + \gamma_1 Z_{i1} A_{it} + \gamma_2 Z_{i2} C_{it} A_{it} + \gamma_3 Z_{i1} Z_{i2} C_{it} A_{it},$$

where $C_{it}$ is a dummy (1/0) indicator of whether or not $Z_{i2}$ was assigned for individual $i$ at time $t$. In the current example $C_{it} = 1\{t > K\}$, where $K$=14 days. This indicator is included to help ensure that the proximal effect of *Prompt* at each decision point is allowed to be impacted only by factors that were introduced prior to that decision point. In the current example, $Z_{i2}$ was assigned on day 14 and hence should be allowed to impact the proximal effect of the prompt only after day 14. Failure to respect this ordering may lead to bias and incorrect conclusions (Dziak et al., 2019; Lu et al., 2016; Nahum-Shani et al., 2020b). Equivalently, $Z_{i2}$ could be set to 0 by convention during the first stage, avoiding the need for the additional notation.

Note that a richer model could be specified to investigate whether the proximal effects are time-varying. For example, linearly time-varying effects could be explored using the model

$$E(Y_{it+\Delta}|\boldsymbol{X}_{i0}, Z_{i1}, Z_{i2}, A_{it}) = \boldsymbol{X}_{i0}\boldsymbol{\beta}_0 + \beta_1 Z_{i1}t + \beta_2 Z_{i2}t^* + \beta_3 Z_{i1}Z_{i2}t^*$$
$$+\gamma_0 A_{it}t + \gamma_1 Z_{i1}A_{it}t + \gamma_2 Z_{i2}A_{it}t^* + \gamma_3 Z_{i1}Z_{i2}A_{it}t^*,$$

where $t$ is time since the randomization of $Z_{i1}$, and $t^*$ is time since the randomization of $Z_{i2}$. Table 7 summarizes how the coefficients in Model (5) can be used to answer a variety of scientific questions about the proximal main effect of the time-varying factor ($A_{it}$) and its interactions with the factors assigned at the beginning of the program ($Z_{i1}$) and at week 2 ($Z_{i2}$). These questions in Tables 6 and 7 are also discussed in Appendix A of Nahum-Shani et al. (2022). Quadratic or other nonlinear effects of time could also be explored.

***Estimation.*** Just as in the hybrid factorial-SMART, some estimands of interest for hybrid SMART-MRT designs require handling the fact that $Z_2$ is not assigned to responders. The W&R approach described earlier in the context of the hybrid factorial-SMART, can also be applied for the hybrid SMART-MRT design (see Nahum-Shani et al., 2022). Recall that this approach involves replicating observations (rows in the dataset) of responders and assigning inverse probability weights to each observation. However, when modeling proximal effects, hybrid SMART-MRT requires a slightly different approach to W&R than hybrid factorial-SMART. This is because the repeated measures of $Y_{it+\Delta}$, one for each $t = 1, \dots, T$, are not independent within individual $i$, so that individual $i$ is a cluster, even before replication. A non-responder would potentially be a cluster of size $T$ (although some occasions may not be counted because of missing data). A responder would be a larger cluster due to the replication of their observations. For example, replicating all data from a responder would lead to a cluster of size $2T$, each row of which would be given half the weight of a data row from a non-responder. However, since $Z_2$ was only assigned on week 2 (day 14) and hence should not have any effect on the proximal outcome before $t = 14$, it should be sufficient to replicate only the data from after the randomization to $Z_2$ and simply treat $Z_2$ as 0 otherwise.

If estimation is being done using standard software, a working independence structure may have to be used for the estimating equations because empirically estimating a $2T \times 2T$ covariance matrix is required in order to calculate the sandwich standard errors for non-working-independence GEE, and this can be computationally impractical. Fortunately, even if a working independence structure is used for parameter estimation, the true dependence structure is accounted for by the sandwich standard error calculation. Thus, the true size of the dataset is not miscounted, and tests still have approximately the appropriate Type One error rate (see simulations in Appendix B of Nahum-Shani et al., 2022).

### Illustrative Data Analysis Based on the Weight Loss Hybrid SMART-MRT

The SMART weight loss study (Pfammatter et al., 2019) is a hybrid SMART-MRT. However, the design structure differs from the structure of the hybrid SMART-MRT in Figure 3 in two important ways. First, the tailoring variable embedded in the SMART was time-varying, meaning that participants' response status was assessed repeatedly, and they were re-randomized to second-stage options at the first time point in which they were classified as non-responders (unless and until that occurred, they continued with the initial intervention option). Second, micro-randomizations were employed only for non-responders and only following their classification as non-responders. Below we describe this study design in more detail and discuss how data from this study can be analyzed to answer scientific questions about possible synergies between components that can be delivered and adapted on different timescales. For illustrative purposes, we generated and analyzed simulated data that mimic key characteristics of real data from this study. These simulated data, as well as the data analysis code and related documentation are available at [https://github.com/d3center-isr/repository-hybrid-trials-QKWHF6125W]. Data were analyzed using R, version 4.2.0 (R Core Team, 2022).

**Study Design**

At the beginning of the study, participants were randomly assigned (with probability 0.5) to either a weight loss smartphone app alone (App) or to an app combined with weekly coaching (App + Coaching). At weeks 2, 4, and 8, participants' early response status was assessed using average weekly weight loss. At the first time point  at which a given participant had an average weekly weight loss of less than 0.5 pounds, they were classified as a non-responder and were randomly assigned (with probability 0.5) to either step-up modestly (by adding text messages) or vigorously (by adding text messages combined with another, more traditional weight loss treatment component that the participant was not offered initially; for details, see Pfammatter et al., 2019) for the remaining intervention period (i.e., until week 12). As long as the participant was classified as a responder, they continued with the initial intervention and were not re-randomized.

The text messages, which were included in both the modest and the vigorous step-up options for non-responders, involved micro-randomizations. Specifically, as soon as the participant was classified as a non-responder, they were randomly assigned daily (in the middle of

The Hybrid Experimental Design

the day), to either a text message that encouraged them to self-monitor their dietary intake (with 0.66 probability) or no message.

**Sample**

The study enrolled 400 individuals between the ages of 18 and 60 with body mass index (BMI) between 27 and 45 kg/m$^2$ (Spring et al., 2020). Given the illustrative nature of this analysis, we simulated data to mimic results from a subsample of 366 individuals (23.8% males; mean age 40.6 years) who did not drop out and who had complete data on all variables required for estimating both the proximal and distal effects discussed below. Of this subsample, 181 participants were randomly assigned to App alone ($Z_1 = -1$), and 185 were randomly assigned to App + Coaching ($Z_1 = 1$) at the beginning of the study. Then, over the twelve-week intervention period, 169 participants were classified as non-responders (96 at the second week, 45 at the fourth week, 28 at the eighth week). Eighty-four non-responders were randomly assigned to the modest step-up and 85 were assigned to the vigorous step-up. 197 participants were classified as responders and continued with their assigned initial intervention.

**Measures**

**Primary proximal outcome.** We consider dietary intake self-monitoring in the next 12 hours as the primary proximal outcome. This is a binary proximal outcome indicating whether (=1) or not (=0) the participant used the mobile app to record their dietary intake in the 12 hours following micro-randomization.

**Primary distal outcome.** We consider weight loss from baseline to month 6 as the distal outcome. Body weight at baseline and month 6 was measured to the nearest 0.25 lb. using a calibrated balance beam scale (see Pfammatter et al., 2019).

**Baseline measures.** Our models included two baseline measures as control covariates: (1) BMI calculated using the Quetelet index as weight in pounds / (height in inches)$^2$ x 704.5; and (2) biological sex.

## Models

This illustrative analysis was motivated by scientific questions concerning possible synergies between three components: (1) offering coaching initially; (2) stepping up vigorously for non-responders; and (3) prompting at mid-day. Specifically, the goal was to investigate (a) whether the proximal effect of prompting (vs. not prompting) on dietary intake self-monitoring in the next 12 hours varies by whether coaching was offered initially (vs. mobile app alone) and by whether a vigorous (vs. modest) step-up was offered to non-responders; and (b) whether the distal effect of offering coaching initially (vs. mobile app alone) on weight loss by month 6 varies by whether a vigorous (vs. modest) step-up was offered to non-responders and by the rate of prompts delivered to non-responders.

Given the unique features of this design, the models for the proximal and distal outcomes had to be modified in several ways. For the proximal outcome, the indicator $C_{it}$ for whether or not $Z_{i2}$ was assigned for individual $i$ at time $t$, was included in all the coefficients involving either $Z_{i2}$ or $A_{it}$ (and not only $Z_{i2}$ as in Model 5), in the following way:

$$\boldsymbol{X}_{i0}\boldsymbol{\beta}_0 + \beta_1 Z_{i1} + \beta_2 C_{it} Z_{i2} + \beta_3 C_{it} Z_{i1} Z_{i2} + \gamma_0 C_{it} A_{it} + \gamma_1 C_{it} Z_{i1} A_{it} + \gamma_2 C_{it} Z_{i2} A_{it} + \gamma_3 Z_{i1} Z_{i2} C_{it} A_{it}.$$

This modification was needed because the daily randomizations to text message vs. no message ($A_{it}$) took place only for non-responders and only after they were classified as non-responders. Further, since the tailoring variable (response status) was time-varying, we included the week in which the participant was classified as a non-responder (this variable was coded 0 for responders) as a control covariate in the model. We also included as a control covariate the number of days elapsed since the start of the subsequent intervention (for responders we considered the end of week 8 as the start of the subsequent intervention). Finally, since the proximal outcome is binary, we used a log-link function such that the effects are expressed on the "risk-ratio" scale. For example, the proximal main effect of delivering a text message (averaging over the other components and all decision points) is expressed in terms of the probability of proximal self-monitoring when a message was delivered, divided by the probability of proximal self-monitoring when a message was not delivered (see Qian et al., 2020).

For the distal outcome model, we investigated the moderating role of the rate of message delivery ($\bar{A}_i$) by restricting the analysis to non-responders ($R = 0$), since only non-responders were randomized daily to message delivery. Specifically, we used the following model:

$$E(Y_i^* | \boldsymbol{X}_{i0}, Z_{i1}, Z_{i2}, \bar{A}_i, R = 0) = \boldsymbol{X}_{i0}\boldsymbol{\theta}_0 + \theta_1 Z_{i1} + \theta_2 Z_{i2} + \theta_3 Z_{i1}Z_{i2}$$
$$+ \theta_4 \bar{A}_i + \theta_5 Z_{i1}\bar{A}_i + \theta_6 Z_{i2}\bar{A}_i + \theta_7 Z_{i1}Z_{i2}\bar{A}_i.$$

**Results**

The results described here are based on simulated data that mimic key characteristics of real data from the SMART weight loss study. Table 8 presents the results for the proximal outcome model. These results indicate that the main effect of delivering (vs. not delivering) a message on proximal self-monitoring of dietary intake, as well as the interactions between this time-varying factor, the initial intervention options, and the subsequent options for non-responders, are not significantly different from zero (i.e., all 95% confidence intervals [CIs] include zero).

Table 9 presents the results for the distal outcome model among non-responders. These results show a significant three-way interaction between the initial options, step-up options, and the rate of messages delivered among non-responders (Est =-14.85; 95% CI: [-24.39, -5.32]). Figure 4 summarizes the estimated weight loss by the initial options, step-up options, and the rate of messages delivered. It shows that weight loss does not vary by the rate of messages delivered for non-responders who started with App alone (regardless of their assigned step-up option). However, non-responders who started with App + Coaching and were then offered a modest step-up, lost more weight to the extent that <u>more</u> messages were delivered. For non-responders who started with App + Coaching and were then offered a vigorous step-up, more weight was lost to the extent that <u>fewer</u> messages were delivered. Overall, these results highlight the possibility that in the context of a mobile-based weight loss intervention that offers a relatively intense level of support (i.e., initial coaching and vigorous step up), high message delivery rate may be too burdensome and hence less effective.

## Discussion

In this paper we presented three types of HEDs. The first integrates a standard factorial design with a SMART design (i.e., hybrid factorial-SMART), the second integrates a standard factorial design with an MRT design (i.e., hybrid factorial-MRT), and the third integrates a SMART design with an MRT design (i.e., hybrid SMART-MRT). For each HED, we have proposed a model and a data analysis method that researchers can employ to answer questions about the selection and adaptation of intervention components on multiple timescales. These designs are especially promising given the increased interest in multimodality interventions that combine human-delivered interventions (e.g., therapy or coaching sessions delivered by a clinician or support group) with digital support (e.g., reminders or motivational messages delivered via a mobile app). Human-delivered and digital components can typically be delivered and adapted on very different timescales, usually slow (e.g., every few weeks or months) in the case of human-delivered components and fast (e.g., every few hours or minutes) in the case of digital components. The HED enables researchers to answer scientific questions about how best to integrate human-delivered and digital support by randomizing participants to human-delivered components and digital components simultaneously at their proper timescales.

This paper is intended not only to serve as a guide for the design and analysis of data from HEDs, but also to serve as a launching point for further research into methodological and practical issues related to these designs. There are multiple ways in which the ideas proposed here can be further developed by additional work. These include the development of sample size planning resources, missing data considerations, extensions to various types of outcomes, and extensions to accommodate variations within each type of HED. Below, we elaborate on each research direction.

Sample size planning for HEDs is an important topic for future research. Power planning resources do exist for standard factorial designs, SMARTs and MRTs. Therefore, a straightforward approach might be to use existing sample size resources for one of the designs comprising the HED; the selected design should likely be the one that is most relevant to the primary scientific question motivating the HED. For example, consider the hybrid SMART-MRT in Figure 3a and 3b. If the primary research question concerns the main effect of the component delivered at program entry (i.e., Add Coaching) in terms of month 6 weight loss (averaging over the other components and all decision points), then investigators may consider planning sample size using existing power resources for SMARTs (Oetting et al., 2007). In this case, power would probably be lower, all else being equal, for the main effect of the component delivered at week 2 (Add Meal) on weight loss at month 6 because not all participants would be randomized on that corresponding factor. Power might be higher for testing the main effect of the time-varying factor *Prompt* on the proximal outcome because this comparison leverages both between-person and within-person contrasts. Alternatively, if the primary question concerns the main effect of the time-varying component Prompt on proximal self-monitoring behaviors (averaging over the other components and all decision points), then investigators may consider planning sample size using existing power resources for MRTs (Liao et al., 2016). In this case, power would probably be lower for testing the main effects of the components at program entry and at week 2. While straightforward, this approach cannot be used to plan sample size for answering scientific questions about some of the interactions between components (e.g., the interaction between Add Coaching and Prompt in terms of the proximal outcome). If sample size for these questions is of interest, it could be investigated with simulations (e.g., see appendix B in Nahum-Shani et al., 2022). Given appropriate simplifying assumptions, sample size formulas might be derived in some cases.

For simplicity of presentation, we did not consider missing data issues in this paper. The simplest way to account for expected missing data when planning a study would be to inflate the planned sample size to compensate (e.g., multiply the planned $n$ by $1/(1-m)$ if a proportion $m$ or less of missingness is expected). When analyzing the data, no guidelines specific to HEDs yet exist. Multiple imputation is probably preferable to listwise deletion, as in other settings, but more work remains to be done on how to best use features of the HED to inform the imputation model. It may also be interesting to study attrition in its own right, perhaps in addition to accounting for it in other analyses. For example, daily prompts might either increase or decrease the likelihood of dropout. Investigators could plan the primary analysis on an intent-to-treat basis and supplement it with an exploratory analysis in which dropout itself is also treated as a binary outcome.

The simple models presented early in this paper were all linear models with an identity link function. This might be inappropriate for some outcome variables, especially binary or categorical outcomes, which could be modeled using a log or logistic link function, as in the empirical data presented in the second part of the paper. Binary or other categorical outcomes would still have implications for power, and perhaps for the specific interpretation of some comparisons, because these link functions generally assume a non-additive model having non-orthogonal effects. For example, floor and ceiling effects or even complete separation can sometimes occur when fitting complicated models to binary data. More work is needed to understand the implications of employing the proposed data analytic methods with different types of outcomes.

In addition to the practical issues described above, the designs presented here can be extended in additional ways. For example, in HEDs that involve the MRT, instead of micro-randomizing participants to the time-varying factor with the same probability at each decision point, randomization probabilities may vary systematically between participants (e.g., in a hybrid factorial-MRT, participants may be randomized at program entry to either be prompted with 0.4 probability on each day or with 0.6 probability on each day). There are many other possible variations depending on the structure of the intervention to be developed and the scientific questions motivating the study. Guidelines are needed for whether and how the proposed data analytic methods should be modified to accommodate design variations.

**Conclusion**

The types of HEDs discussed here are all extensions of factorial designs. They share some of the benefits of factorial designs, such as the ability to explore interactions and the efficiency gained from investigating multiple factors in a single sample. They may also share the challenges of factorial designs, such as the need for appropriate model assumptions, the task of prioritizing the most important questions out of the many which can be answered, and the importance of planning for feasibility and fidelity of treatment delivery in multiple different conditions. Further work will help address these challenges as HEDs become more prevalent in behavioral and health sciences.

Table 1: Adaptive interventions embedded in the hybrid factorial-SMART in Figure 1

| Embedded adaptive intervention | At program entry | At week 2 | | Cells |
|---|---|---|---|---|
| 1 | App and Coaching | Responders | Continue | 1→A |
| | | Non-responders | Add Meal | 1→B |
| 2 | App and Coaching | Responders | Continue | 1→A |
| | | Non-responders | | 1→C |
| 3 | App alone | Responders | Continue | 2→A |
| | | Non-responders | Add Meal | 2→B |
| 4 | App alone | Responders | Continue | 2→A |
| | | Non-responders | | 2→C |
| 5 | Coaching alone | Responders | Continue | 3→A |
| | | Non-responders | Add Meal | 3→B |
| 6 | Coaching alone | Responders | Continue | 3→A |
| | | Non-responders | | 3→C |
| 7 | No App and No Coaching | Responders | Continue | 4→A |
| | | Non-responders | Add Meal | 4→B |
| 8 | No App and No Coaching | Responders | Continue | 4→A |
| | | Non-responders | | 4→C |

Table 2: Scientific questions and model parameters for a hybrid factorial-SMART

| | Scientific question | Type | Model (1) parameters |
|---|---|---|---|
| A | Is it beneficial (in terms of month 6 weight loss) to offer (or not to offer) the App at the beginning of the program? | Main effect of initial factor, averaging over the other factors and response status | $2\theta_1$ |
| B | Is it beneficial (in terms of month 6 weight loss) to offer (or not to offer) Coaching at the beginning of the program? | Main effect of initial factor, averaging over the other factors and response status | $2\theta_2$ |
| C | Is it beneficial (in terms of month 6 weight loss) to offer (or not to offer) Meal at week 2 for non-responders? | Main effect of subsequent factor for non-responders, averaging over the other factors and response status | $2\theta_4$ |
| D | Is the App more beneficial (in terms of month 6 weight loss) with or without Coaching? | Two-way interaction between the two initial factors, averaging over the subsequent factor and response status | $2\theta_3$ |
| E | Is the App more beneficial (in terms of month 6 weight loss) with or without offering Meal to non-responders at week 2? | Two-way interaction between initial factor and subsequent factor for non-responders, averaging over the other initial factor and response status | $2\theta_5$ |
| F | Is Coaching more beneficial (in terms of month 6 weight loss) with or without offering Meal to non-responders at week 2? | Two-way interaction between initial factor and subsequent factor for non-responders, averaging over the other initial factor and response status | $2\theta_6$ |
| G | Does the interaction between *App* and *Coaching* vary depending on whether or not Meal is offered to non-responders? | Three-way interaction between the two initial factors and the subsequent factor for non-responders, averaging over response status | $2\theta_7$ |

Table 3: Scientific questions about distal effects and model parameters for a hybrid factorial-MRT

| | Scientific question about distal effects | Type | Model (1) parameters |
|---|---|---|---|
| A | Is it beneficial (in terms of month 6 weight loss) to add (or not to add) coaching at the beginning of the program? | Main effect of factor assigned at program entry, averaging the other factor assigned at program entry and the time-varying factor | $2\theta_1$ |
| B | Is it beneficial (in terms of month 6 weight loss) to add (or not to add) meal replacement at the beginning of the program? | Main effect of factor assigned at program entry, averaging the other factor assigned at program entry and the time-varying factor | $2\theta_2$ |
| C | Is adding coaching more beneficial (in terms of month 6 weight loss) with or without adding meal at the beginning of the program? | Two-way interaction between two factors that were assigned at program entry, averaging over the time-varying factor | $2\theta_3$ |
| D | Does the effect of (i.e., difference between) adding (vs. not adding) coaching at the beginning of the program vary by the rate of prompt delivery? | Two-way interaction between a factor assigned at program entry and the rate of the time-varying factor, averaging over the other factor assigned at program entry. | $\theta_5(\bar{a} - \bar{a}')$ |
| E | Does the effect of (i.e., difference between) adding (vs. not adding) meal replacement at the beginning of the program vary by the rate of prompt delivery? | Two-way interaction between a factor assigned at program entry and the rate of the time-varying factor, averaging over the other factor assigned at program entry. | $\theta_6(\bar{a} - \bar{a}')$ |
| F | Does the interaction between *Add Coaching* and *Add Meal* varies by the rate of prompt delivery? | Three-way interaction between factors assigned at program entry and the time-varying factor | $\theta_7(\bar{a} - \bar{a}')$ |

Table 4: Scientific questions about proximal effects and model parameters for a hybrid factorial-MRT

| | Scientific question about proximal effects | Type | Model (3) parameters |
|---|---|---|---|
| A | Is it beneficial (in terms of increasing self-monitoring of dietary intake in the next 12 hours) to deliver (vs. not deliver) a mobile-based prompt every day at mid-day? | Main effect of the time-varying factor, averaging over the factors that were assigned at program entry and all decision points | $2\gamma_0$ |
| B | Does the effect of delivering (vs. not delivering) a daily prompt on self-monitoring of dietary intake in the next 12 hours varies by whether or not coaching was initiated at the beginning of the program? | Two-way interaction between the time-varying factor and one of the factors that was assigned at program entry, averaging over the other factor and all decision points | $2\gamma_1$ |
| C | Does the effect of delivering (vs. not delivering) a daily prompt on self-monitoring of dietary intake in the next 12 hours varies by whether or not meal replacement was initiated at the beginning of the program? | Two-way interaction between the time-varying factor and one of the factors that was assigned at program entry, averaging over the other factor and all decision points | $2\gamma_2$ |
| D | Does the effect of delivering (vs. not delivering) a daily prompt on self-monitoring of dietary intake in the next 12 hours vary by the interplay between *Add Coaching* and *Add Meal*, averaging over all decision points? | Three-way interaction between the time-varying factor and the two factors assigned at program entry | $2\gamma_3$ |

Table 5: Adaptive interventions embedded in the hybrid SMART-MRT in Figure 3

| Embedded adaptive intervention | At program entry | At week 2 | | Cells |
|---|---|---|---|---|
| 1 | App alone | Responders | Continue | 1→A |
| | | Non-Responders | Add Meal | 1→B |
| 2 | App alone | Responders | Continue | 1→A |
| | | Non-Responders | | 1→C |
| 3 | App and Coaching | Responders | Continue | 2→A |
| | | Non-Responders | Add Meal | 2→B |
| 4 | App and Coaching | Responders | Continue | 2→A |
| | | Non-Responders | | 2→C |

The Hybrid Experimental Design

Table 6: Scientific questions about distal effects and model parameters for a hybrid SMART-MRT

| | Scientific question about distal effects | Type | Model (4) parameters |
|---|---|---|---|
| A | Does it benefit month 6 weight loss to add coaching to the app at the beginning of the program? | Main effect of factor assigned at program entry, averaging over the factor assigned at week 2 for non-responders and the time-varying factor | $2\theta_1$ |
| B | Does it benefit month 6 weight loss to add meal replacement at week 2 for non-responders? | Main effect of factor assigned at week 2 for non-responders, averaging over the factor assigned at program entry and the time-varying factor | $2\theta_2$ |
| C | Does the beneficial effect (on 6 month weight loss) of adding meal replacement for non-responders at week 2 depend on whether or not coaching was initiated at program entry? | Two-way interaction between the factor assigned at program entry and the factor assigned at week 2 for non-responders, averaging over the time-varying factor | $2\theta_3$ |

Table 7: Scientific questions about proximal effects and model parameters for a hybrid SMART-MRT

| | Scientific question about proximal effects | Type | Model (5) parameters |
|---|---|---|---|
| A | Does delivering a prompt improve dietary self-monitoring in the next 12 hours? | Main effect of the time-varying factor, averaging over the other factors and all decision points | $2\gamma_0$ |
| B | Does the effect of delivering a prompt (on self-monitoring in the next 12 hours) depend on whether coaching was initiated at the beginning of the program? | Two-way interaction between the time-varying factor and the factor assigned at program-entry, averaging over the other factor and all decision points | $2\gamma_1$ |
| C | Does the effect of delivering a prompt (on self-monitoring in the next 12 hours) depend on whether meal replacement was offered to non-responders at week 2? | Two-way interaction between the time-varying factor and the factor assigned at week 2 for non-responders, averaging over the other factor and all decision points | $2\gamma_2$ |
| D | Does the *extent to which* the effect of delivering a prompt (on self-monitoring in the next 12 hours) depends on the delivery of meal replacement to non-responders at week 2, depend itself on whether coaching was initiated at program entry? | Three-way interaction between the time-varying factor, the program-entry factor and the week 2 factor for non-responders | $2\gamma_3$ |

Table 8: Proximal model results

| | Parameter | Estimate | *SE* | 95% CI LL | 95% CI UL |
|---|---|---|---|---|---|
| Control covariates* | $Z_{i1}$: App+Coaching (= 1) vs. App alone (= -1) | 0.04 | 0.0040 | 0.03 | 0.05 |
| | $Z_{i2}$: Vigorous (= 1) vs. Modest (= -1) augmentation | 0.00 | 0.0053 | -0.01 | 0.01 |
| | Week classified as a non-responder | -0.02 | 0.0015 | -0.02 | -0.01 |
| | Biological sex (Female= 1; Male= -1; mean centered) | 0.01 | 0.0046 | 0.00 | 0.02 |
| | Days since classified as a non-responder | -0.00 | 0.0002 | -0.00 | 0.00 |
| | Baseline BMI (mean centered) | -0.01 | 0.0006 | -0.01 | -0.00 |
| Causal effects | $A_{it}$: Message (= 1) vs. No Message (= -1) | 0.01 | 0.01 | -0.007 | 0.026 |
| | $A_{it}Z_{i1}$ | -0.01 | 0.01 | -0.026 | 0.001 |
| | $A_{it}Z_{i2}$ | 0.01 | 0.02 | -0.023 | 0.049 |
| | $A_{it}Z_{i1}Z_{i2}$ | 0.01 | 0.02 | -0.030 | 0.053 |

*Notes*:

Proximal outcome: self-monitoring of dietary intake (yes/no) in the next 12 hours

SE: standard error; CI: confidence interval; LL: lower limit; UL: upper limit; BMI: body mass index

*Although estimates pertaining to the control variables are provided for completeness, we caution readers against interpreting them since correct specification of this part of the model is not required to yield consistent estimates of the causal effect of the randomized messages (see Boruvka et al., 2017).

Table 9: Distal model results for non-responders, investigating the role of message rate

| Parameter | Estimate | *SE* | 95% CI LL | 95% CI UL |
|---|---|---|---|---|
| Intercept | 3.76 | 0.54 | 2.71 | 4.81 |
| Biological sex (Female=1; Male=-1) | 0.88 | 0.69 | -0.47 | 2.23 |
| Baseline BMI (mean centered) | -0.23 | 0.08 | -0.39 | -0.07 |
| $Z_{i1}$: App+Coaching (=1) vs. App alone | 1.88 | 0.54 | 0.83 | 2.94 |
| $Z_{i2}$: Vigorous (=1) vs. Modest (=-1) augmentation | 0.25 | 0.54 | -0.81 | 1.31 |
| $\bar{A}_i$: Rate of message delivery (mean centered) | 7.12 | 4.89 | -2.46 | 16.69 |
| $Z_{i1}Z_{i2}$ | 0.12 | 0.53 | -0.93 | 1.16 |
| $Z_{i1}\bar{A}_i$ | 6.03 | 4.86 | -3.50 | 15.55 |
| $Z_{i2}\bar{A}_i$ | -11.57 | 4.86 | -21.08 | -2.05 |
| $Z_{i1}Z_{i2}\bar{A}_i$ | -14.85 | 4.86 | -24.39 | -5.32 |

Distal outcome: weight loss by month 6

SE: standard error; CI: confidence interval; LL: lower limit; UL: upper limit; BMI: body mass index

The Hybrid Experimental Design

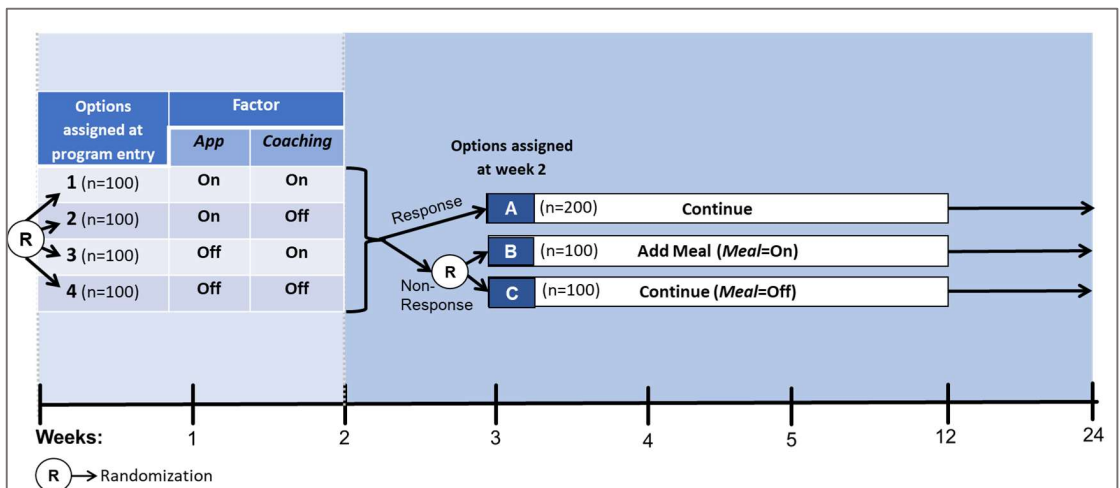

**Figure 1a: An Example Hybrid Factorial-SMART**

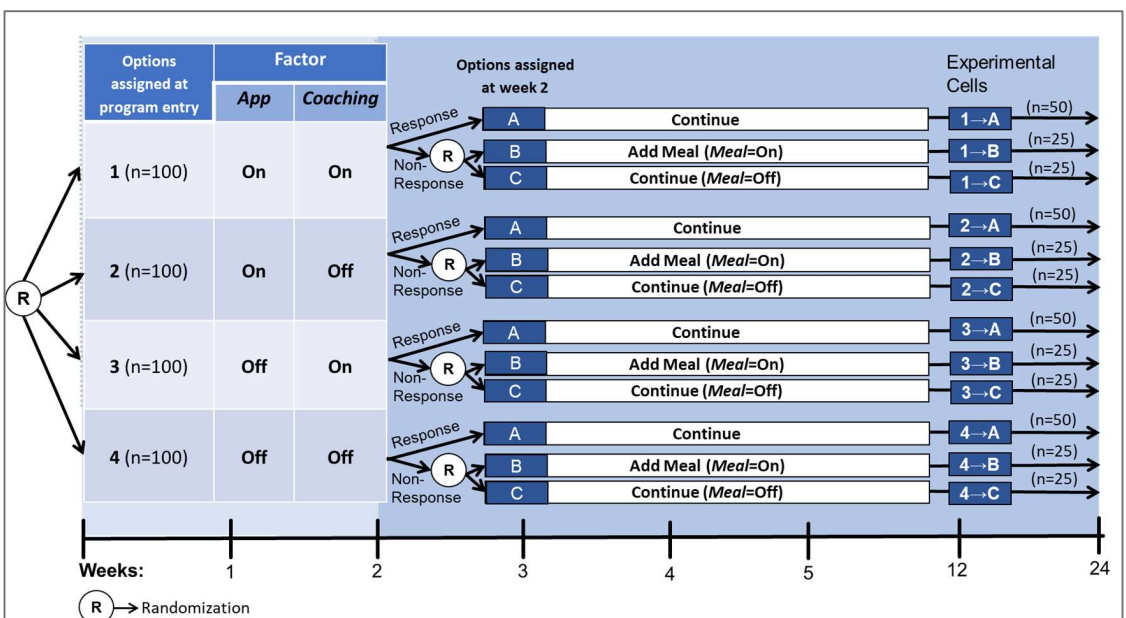

**Figure 1b: Experimental Cells in The Example Hybrid Factorial-SMART**

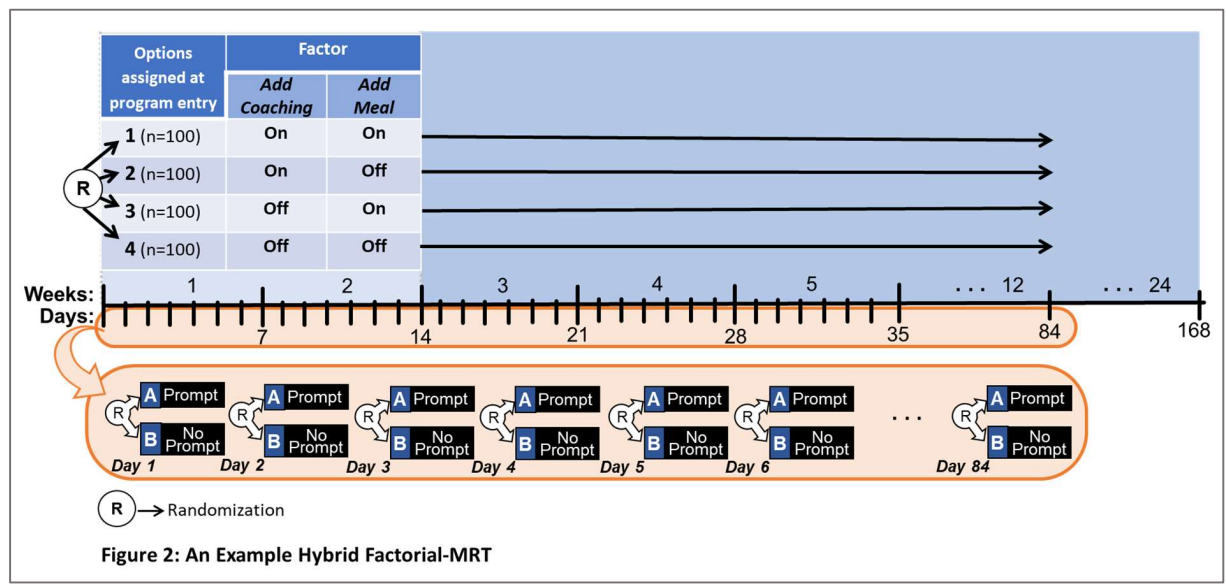

**Figure 2: An Example Hybrid Factorial-MRT**

The Hybrid Experimental Design

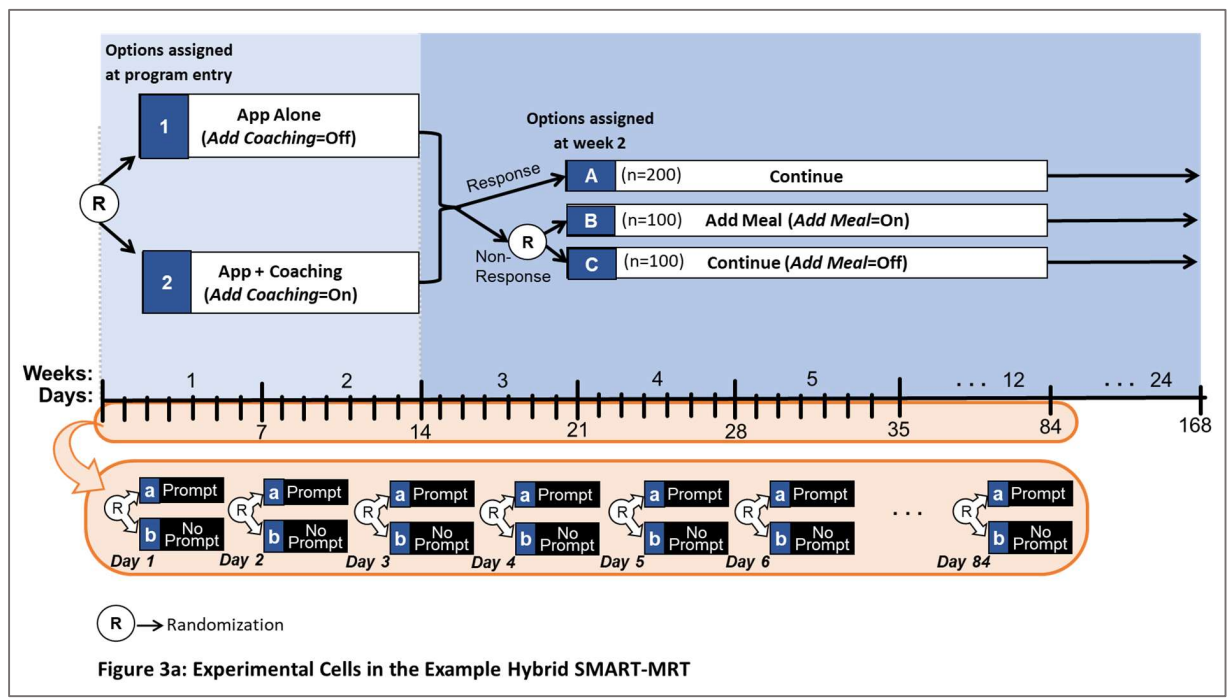

**Figure 3a: Experimental Cells in the Example Hybrid SMART-MRT**

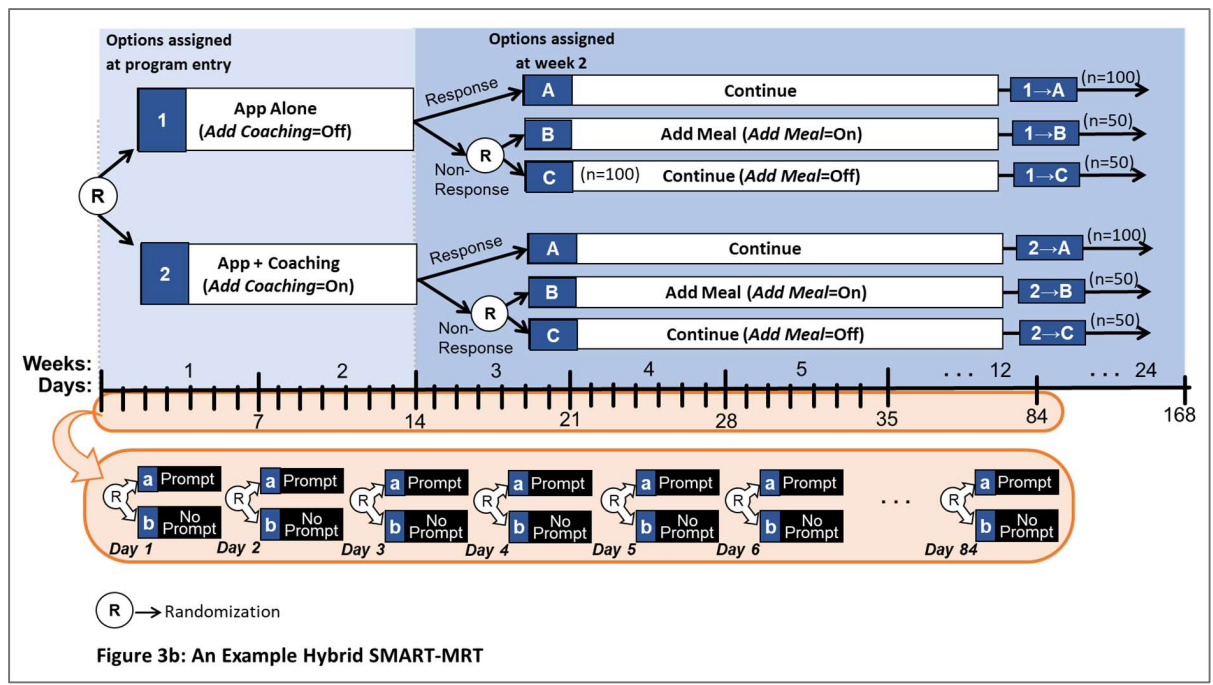

**Figure 3b: An Example Hybrid SMART-MRT**

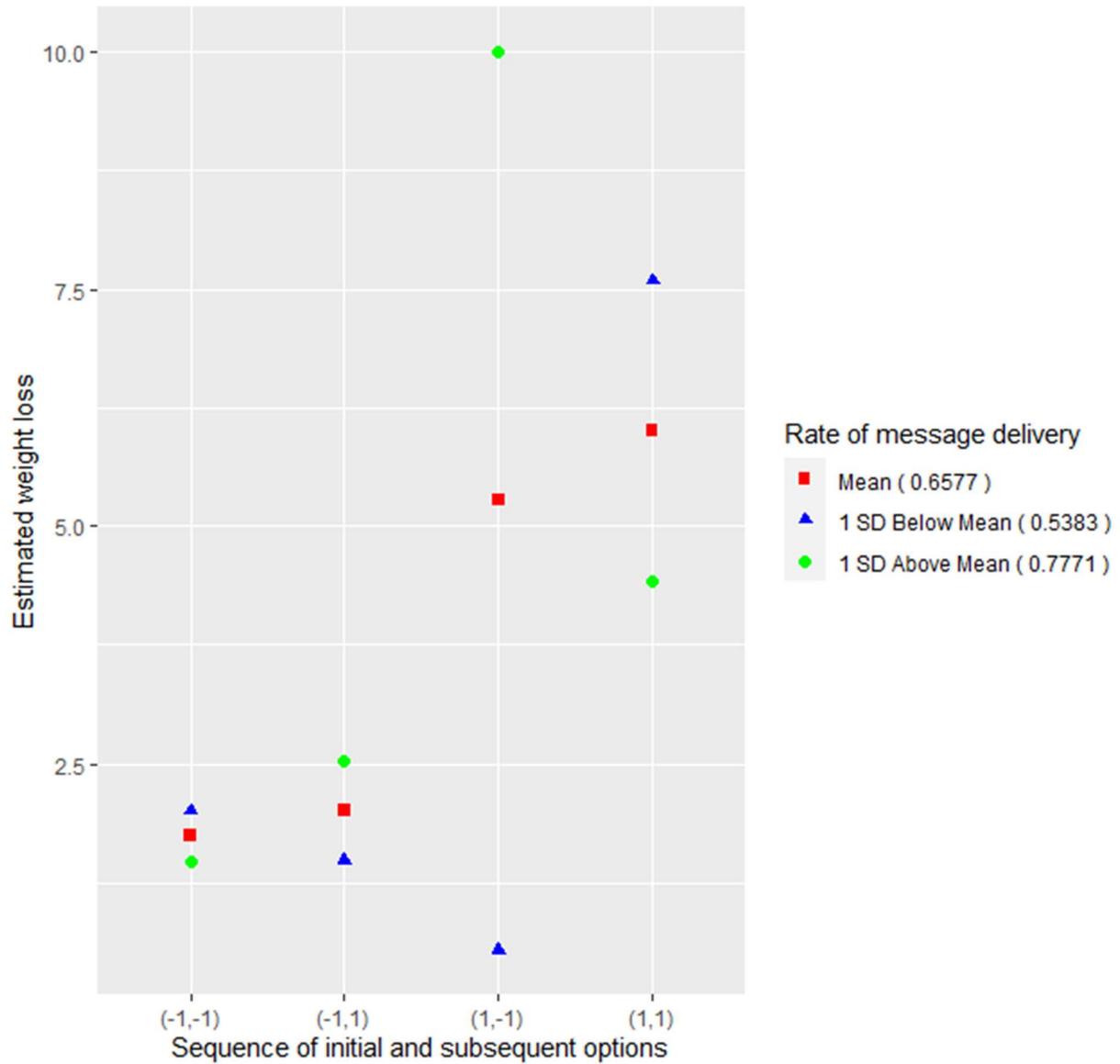

Figure 4: Estimated weight loss by the initial options, subsequent options and rate of message delivery